\documentclass[aps,pre,twocolumn,showpacs]{revtex4}
\usepackage{amsmath}
\usepackage{amssymb}
\usepackage[pdftex]{graphicx}
\usepackage{natbib}
\usepackage{color}
\usepackage{epsfig}
\usepackage{epstopdf}
\usepackage{mhchem}
\usepackage{times,mathptmx}

\begin{document}

\bibliographystyle{apsrev}

\title{Peristaltic pumps work in nano scales}

\author{F. Farahpour}
\affiliation{Sharif University of Technology, Department of Physics, P.O. Box 11155-9161, Tehran, Iran. Fax: +98 21 66022711; Tel: +98 21 66164501}

\author{M.R. Ejtehadi}
\email{ejtehadi@sharif.edu}
\affiliation{Sharif University of Technology, Department of Physics, P.O. Box 11155-9161, Tehran, Iran. Fax: +98 21 66022711; Tel: +98 21 66164501}

\date{\today}

\begin{abstract}
A design for a pump is suggested which is based on well known peristaltic pumps. In order to simply describe the operation of the proposed pump, an innovative interpretation of low Reynolds number swimmers is presented and thereafter a similar theoretical model would be suggested to quantify the behavior of the pumps. A coarse-grained molecular dynamic simulation is used to examine the theoretical predictions and measure the efficiency of the pump in nano scales. It is shown that this pump with a modest design is capable of being a good option for transport processes in nano scale.

\end{abstract}

\pacs{Valid PACS appear here}

\maketitle
\newcommand{\myepsilon}{\text{\usefont{OML}{cmr}{m}{n}\symbol{15}}}
\section{Introduction}
During the last decades there was an increasing interest in fabrication of systems in nano and micro scales and according to the importance of pumps as a part of microfluidic systems, so far great efforts have been done to fabricate pumps in miniature scales.
During recent years several different kind of micro and nano pumps, based on different pumping methods and principles have been designed
and constructed \cite{Laser2004,Woias2005,Tsai2007,Iverson2008,Chen2010,Wang2012}.
\\
Beside all of these efforts, fabrication and developing of valveless micropumps attracted more attention. Having no valve in fact means
the elimination of some of mechanical parts of micropump which make the whole pump, ineffective in short time because of their unremitting wear and fatigue. One of the most important category of valveless micropumps, which in last years some different kind of it was manufactured and developed, is peristaltic pumps. Peristaltic and Pneumatic micropumps using piezoelectric disks as actuators, are one of the first developed micropumps \cite{Olsson1998}. Simply one can consider a peristaltic pump as a long tube in which contractile waves traveling along it, induce a fluid flow inside it \cite{Berg2003,Laser2004}. Characteristics and efficiency of these simple pumps are vastly investigated in literature and is used for micro pumping in biology and industry \cite{Lee2012a,Goldschmidtboing2005,Chang2007,Knight2004,Jeong2007,Huang2006,Berg2003}. Simplicity of the structure of peristaltic and Pneumatic pumps is a great advantage which can be a base for their miniaturization. 
\\
Fabrication of nanofluidic structures is more challenging than microfluidic systems. This is in principal due to difficulty of manufacturing of nano scale structures and significant changes in fluids behavior because of changing the size of the system in several orders of magnitude. Nevertheless, during last years some nanopumps have been designed and developed \cite{Wang2012,Wang2007,Xu2012a,Lohrasebi2012,Chen2010,Tuzun1996,Zhang2008}.
\\
In 2003, Berg \textit{etal.} suggested a two-stage discrete peristaltic micropump which seems to be the simplest kind of peristaltic pumps \cite{Berg2003}. This pump is only consisted of three connector tubes and two actuators which their specific periodic motion yields to the fluid flow in a distinctive direction. This pump needs no valve and in addition it is a bi-directional pump, which means that only by reversing the order of operation of two actuators, the flow direction will be reversed. 
\\
One of the most important benefits of this pump which would be discussed in this paper is its capability to be miniaturized to nanometer scale. In addition to its structural simplicity, the actuation sequence in one full pumping cycle of this pump is non$-$reciprocal. One of the known challenges in micro and nano scales (low Reynolds regime) to create an effective directional motion or flow is that exertion of any force to the fluid in this regime could be completely neutralized by a time-reversed force \cite{Nelson} but it has been shown that a non reciprocal
periodic motion generally can yield to an effective motion \cite{Shapere1989,Nelson}.
\\
In some feature, Operation of this pump is similar to non-reciprocal swimmers have already been introduced \cite{Purcell1977,Najafi2004,Avron2005}. In the next section of this paper this similarity would be described more clearly.
\\
In Berg's suggested pump, actuator regions should be opened and closed completely but one can show that even in half-closed state, this pump
can still generate a directional flow but its efficiency depends on the details of the procedure such as difference between actuators' volume in open and close states and frequency of actuation. A simple analogous model for the Berg's pump can be considered in which whole pump is modeled as a long tube which connects two reservoir tanks. Two regions of the tube, are flexible and act as actuators. In the next section this model would be discussed in detail. 
\section{Analogy between Berg pump and low Reynolds micro swimmers}
\begin{figure}
\begin{center}
\begin{tabular}{c}
\includegraphics[width=.8\columnwidth]{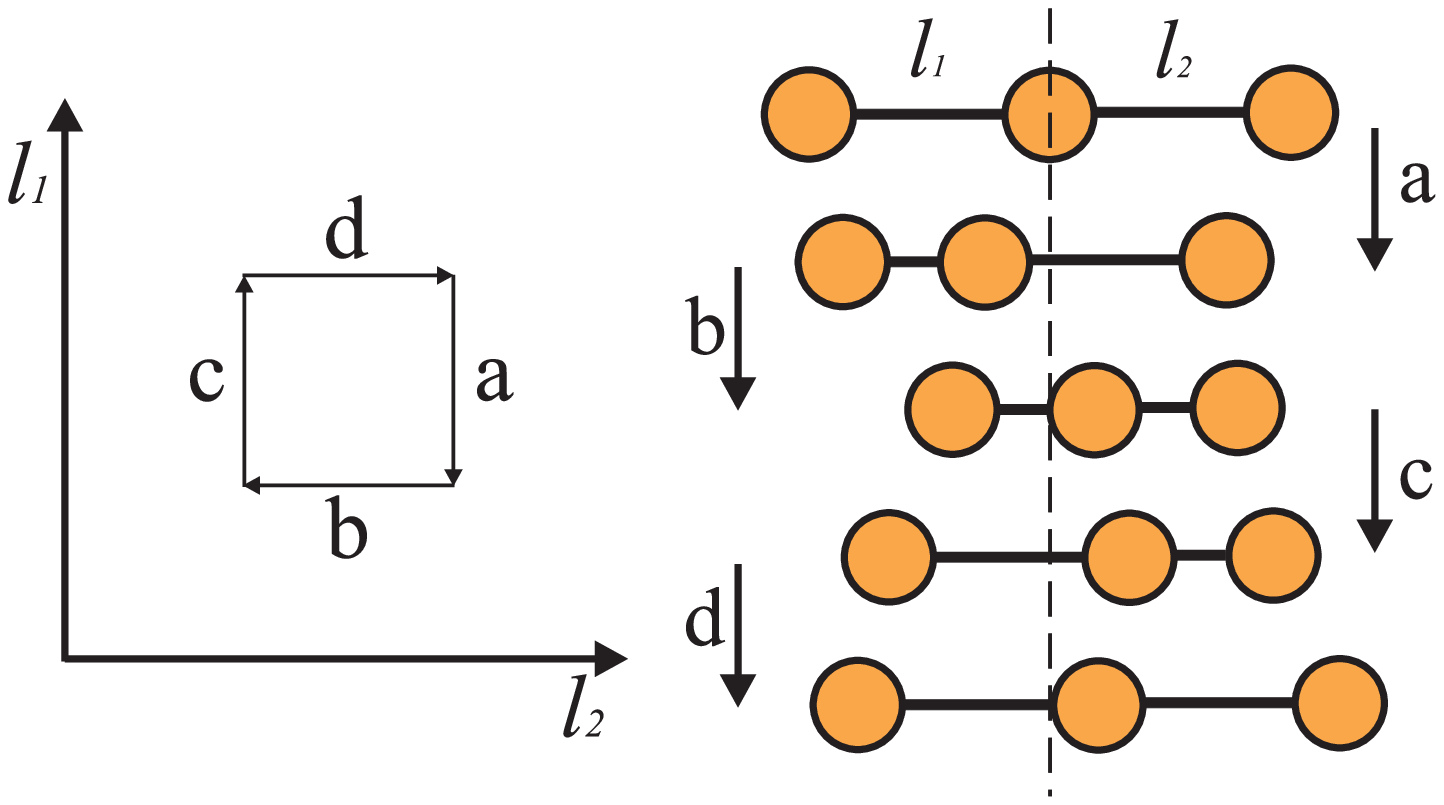} \\
$(a)$\\
\includegraphics[width=.8\columnwidth]{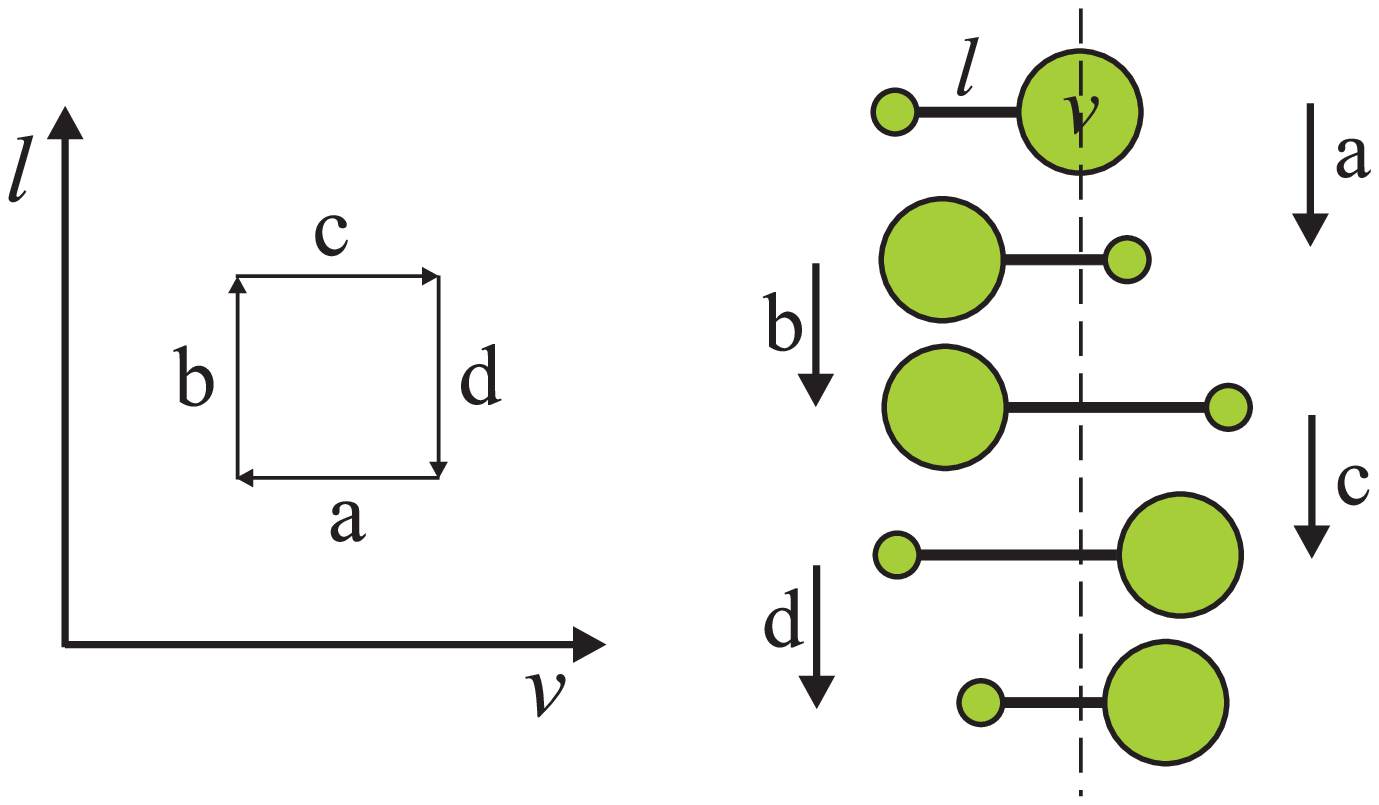} \\ 
$(b)$ \\
\end{tabular}
\end{center} \caption{Operation of two micro swimmers and their non-reciprocal loops. a) NG swimmer, with three fixed-volume spheres and two variable-length bonds. b) AKO swimmer, two variable-volume spheres and one variable-length bond.
\label{Golestanian-Avron}}
\end{figure}

\begin{figure}
\begin{center}
\includegraphics[width=.9\columnwidth]{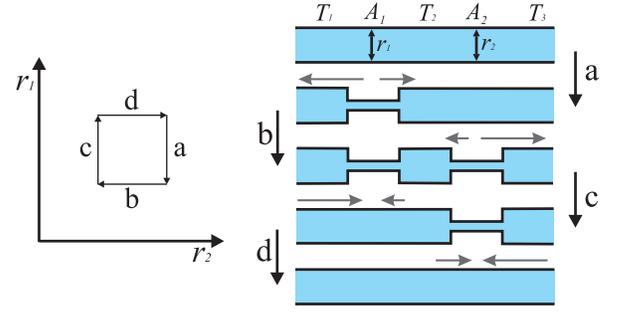}
\end{center} \caption{Operation of Berg pump and its non-reciprocal loop in $r_1-r_2$ space. Flow direction and magnitude in each step is shown schematically with gray arrows.\label{nanopump}}
\end{figure}

\begin{figure}
\begin{center}
\includegraphics[width=.4\columnwidth]{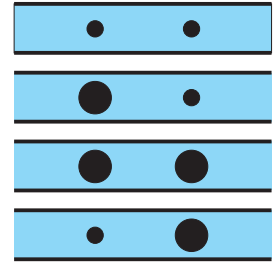}
\end{center} \caption{The simple model used in simulation for actuation of pump. \label{tube-bead}}
\end{figure}

For better understanding of operation of berg pump in nano scale, one can establish a similarity between Berg pump and low Reynolds micro swimmers like the swimmer introduced by Najafi and Golestanian (NG) or the one designed by Avron, Kenneth and Oaknin (AKO) which are the simplest non-reciprocal swimmers and their motion can be described just by two independent parameters. one can shows that their motion in phase$-$space, constructed by these parameters, forms a closed loop in a full cycle, instead of a repetitive back and forth motion. For example NG swimmer can be described by the length of its two arms. This model swimmer consists of three spheres with radius $R$ that are connected by two parallel rigid slender arms, with lengths $l_1$ and $l_2$. There are two internal engines on the middle sphere, which act as internal active elements responsible for making a nonreciprocal motion which is needed to propel the whole system \cite{Najafi2004}. One complete period of its motion, as can be seen in Fig. \ref{Golestanian-Avron}-a, could be represented by a closed loop in $l_1-l_2$ space. It is shown that this swimmer by breaking the time-reversal symmetry as well as the translational symmetry can swim at low Reynolds number. Similarly, AKO swimmer, which has a non-reciprocal motion in $l-v$ space (Fig. \ref{Golestanian-Avron}-b), is another example of such swimmers \cite{Avron2005}.
\\
In an innovative look to NG swimmer, with a simple analysis one can realize that this swimmer, in fact, in a non-reciprocal cycle in
$l_1-l_2$ space, changes its drag coefficient in such a way that forces can not be eliminated by each other in forward and backward motion. This changes in drag coefficient is exactly the origin of the breaking in time-reversal symmetry and hence the effective displacement of the swimmer. To show the simplicity of this analysis we assume that drag coefficient of two adjacent spheres, when they are far away from each other (expansion of the connector arm), is equal to ${\zeta }_1$ but when two beads are close to each other (contraction of the connector arm) the drag coefficient of their collection is ${\zeta }_2<2{\zeta }_1$, due to hydrodynamic interaction between them \cite{Nelson}. Meanwhile we consider that swimmer's arms expand and contract by a uniform and constant velocity in whole cycle. The length of the arms in expanded and contracted state is $L$ and $L-e$, respectively. Periodic motion of this swimmer, as shown in Fig. \ref{Golestanian-Avron}-a, is consisted of left-arm contraction, right-arm contraction, left-arm expansion and finally, right-arm expansion.
\\
By considering the stokes' law for the motion in viscous environments, one can calculate the magnitude of displacement of the middle bead.
In the $i$th stage of the motion, $i=1,\dots,4$, the velocity of the middle bead relative to the fluid, which is considered stationary, is $v_i$
and its displacement is $\Delta x_i$. Thus, for example, in the first stage,
\begin{eqnarray}\
{\zeta }_1\left(u+v_1\right)+{\zeta }_1v_1+{\zeta }_1v_1=0\
\end{eqnarray}
is the governing equation of the swimmer's motion and therefore $v_1=-\frac{u}{3}\ $and $\Delta x_1=-\frac{e}{3}$. But in the second stage, the drag coefficient of the constituents of the swimmer will change as described above and thus governing equation will change to a new form 
\begin{eqnarray}\
{\zeta }_2v_2+{\zeta }_1\left(v_2-u\right)=0\
\end{eqnarray}
and $v_2=\frac{{\zeta }_1}{{\zeta }_1+{\zeta }_2}u$ and $\Delta x_2=\frac{{\zeta }_1}{{\zeta }_1+{\zeta }_2}e$. Repeating these calculations in the next stages, gives us $\Delta x_3=\frac{{\zeta}_1}{{\zeta }_1+{\zeta }_2}e$ and $\Delta x_4=-\frac{e}{3}$, so in a complete cycle of motion, forward displacement of the swimmer is obtained by
\begin{eqnarray}\
\Delta x=\sum_i{\Delta x_i{\rm =}\frac{{\rm 2}}{{\rm 3}}{\rm
e(}\frac{{\rm 2}{\zeta }_{{\rm 1}}{\rm -}{\zeta }_{{\rm 2}}}{{\zeta
}_{{\rm 1}}{\rm +}{\zeta }_{{\rm 2}}}{\rm )}>0}.
\end{eqnarray}
\\
In order to make an analogy between Berg pump and NG swimmer one can consider a simple model for the Berg pump in which whole pump is modeled as a long tube which connects two reservoir tanks. Two regions of the tube, are flexible and act as actuators (Fig.\ref{nanopump}). 
\\
For example one can assume three carbon nano tubes, which are connected together via two protein nano-pores, like OmpF channel, a biological pore found in the outer membrane of Escherichia coli, and the width of these nano-pores can be changed by a specific activation in a controlled manner. Another model which can be considered is a long tube that two specific proteins, with two activation modes (contracted and expanded), or thermosensitive particles, with hydrodynamic radius dependent on temperature, are fixed in two regions of it (e.g. attached to its interior wall). By applying an AC electromangnetic radiation or focusing laser beams on their positions, \cite{Senff1999} one may be able to activate the proteins in such a way that widening and narrowing of two actuator parts of the Berg's pump is modeled. Such mechanisms of contraction and expansion of proteins in nanoscale is known today for some biologic gates, like OmpF, in selective transferring of ions. 
\\
Two flexible actuator regions of the tube are indexed by $A_1$ and $A_2$ (Fig.\ref{nanopump}). Three connector tubes are $T_1$, $T_2$ and $T_3$. The width of $A_1$ and $A_2$ regions decreases and increases in the order which is shown in Fig. \ref{nanopump}, hence an effective flow is generated. The characteristic parameters of this pump are the radius of $A_1$ and $A_2$ regions ($r_1$ and $r_2$) which their variation in $r_1-r_2$ space forms a nonreciprocal loop. For simplicity we consider that the changes in volume of actuators are made by varying their radius in a uniform way, and the radius of an actuator is equal throughout its length. The real deformation of actuator is certainly smoother, but this assumption has no effect in generality of the pump's operation.
\\
In Berg pump, resistance of two flexible regions changes exactly in the same way of the drag coefficient in NG swimmer. The fluid flow inside a pipe in low Reynolds regime can be obtained from a simple relation, $Q=\frac{p}{Z}$ \cite{Nelson}, in which $p$ is the pressure difference between two ends of the pipe and $Z=\frac{8L\eta }{\pi R^4}$ is the pipe resistance against fluid flow. The dependence of resistance on the forth power of pipe's radius, gives us this ability to make an effective change in the resistance of the pipe just by a little change in its radius. We assume that narrowing a region of the pipe increases the pressure in that region from $p$ to $p+\Delta p$ and widening it decreases the pressure to $p-\Delta p$. The length of $A_1$ and $A_2$ regions and $T_1$, $T_2$ and $T_3$ connector tubes are $L$, initial radius of whole tube is $R$ and the radius of flexible parts in contracted state is $r$. In addition we suppose that pressure is uniform in a tube by passing a specific time after a local variation. There is a delay time between two consecutive stages, $\delta t$, which is essential for this last condition to be established and finally we define the resistance of the tubes in expanded and contracted form as $Z_e=\frac{8L\eta }{\pi R^4}$ and $Z_c=\frac{8L\eta }{\pi r^4}$, respectively.
\\
In the first stage of motion, pressure in $A_1$ region increases to $p+\Delta p$. Flow resistance for right and left part of the first actuator in this stage are $Z_r=\frac{8(3L)\eta}{\pi R^4}=3Z_e$ and $Z_l=\frac{8L\eta }{\pi R^4}=Z_e$, respectively. $Z_r$ is obtained by adding the resistance of two connector tubes ($T_2$ and $T_3$) with the right actuator ($A_2$) in expanded state but $Z_l$ is only resistance of one connector tube ($T_1$). So resistance against right direction flow in right part is greater and flow in right and left part will be $Q_{1r}=+\frac{\Delta p}{{3Z}_e}$ and $Q_{1l}=-\frac{\Delta p}{Z_e}$ (Gray arrows in Fig.\ref{nanopump}). 
\\
In second stage, pressure in $A_2$ actuator increases but this time $A_1$ is half-closed and therefore $Z_r=Z_e$ (resistance of $T_3$ connector tube in the right side of $A_2$ actuator) and $Z_l=2Z_e+Z_c$ (sum of resistance of two connector tubes ($T_1$ and $T_2$) and the left actuator ($A_1$) in contracted state), so for flows, one obtaines $Q_{2r}=+\frac{\Delta p}{Z_e}$\textit{ }and $Q_{2l}=-\frac{\Delta p}{2Z_e+Z_c}$.
\\
In the third and forth stages, after widening the actuators sequentially the pressure decreases in respective actuator and fluid flows
toward the operating actuator (Fig.\ref{nanopump}). After some straightforward calculation one reaches to $Q_{3r}=-\frac{\Delta p}{2Z_e+Z_c}$ ,$Q_{3l}=+\frac{\Delta p}{Z_e}$ ,$Q_{4r}=-\frac{\Delta p}{Z_e}$ and $Q_{4l}=+\frac{\Delta p}{3Z_e}$. Finally, after summing the flow magnitude in all of the stages, one reaches to the total amount of flow in the whole cycle:
\begin{eqnarray}\
Q=\sum_i{\left(Q_{ir}+Q_{il}\right)=+2\Delta
p(\frac{1}{3Z_o}}-\frac{{\rm 1}}{2Z_o+Z_e}).\
\end{eqnarray}
\\
According to the definition of $Z_e$ and $Z_c$ and by considering the fact that $r<R$, one can see that the expression in the parentheses is always positive and hence there is a fluid flow in right direction inside the pipe. In reverse order, when the second actuator is closed at first, a leftward flow is generated.
\\
To show the effective operation of this pump in nano scale, we constructed a simple model and investigated its general performance based on some relevant parameters of the system by a molecular dynamic simulation. To simplify the process of modeling and without any loss of generality, we modeled the pump with a long connecting pipe between input and output reservoir and the actuators by putting two fixed beads with variable volume in each actuator part. The volume of these beads increases and decreases according to the algorithm mentioned above (in order to make a contraction in each actuator, the corresponding bead should expand and vice versa) (Fig. \ref{tube-bead}). Such a modeling has a great advantage which can show the duality between swimmers and pumps more distinctly. In other words one can imagine that this pump is a swimmer consisted of a rod with fix length that in two ends is connected to two beads with variable volume and the
swimmer as a whole is fixed in middle point of the rod.
\section{Simulations and results}
\begin{figure}
\begin{center}
\begin{tabular}{cc}
\includegraphics[width=.48\columnwidth]{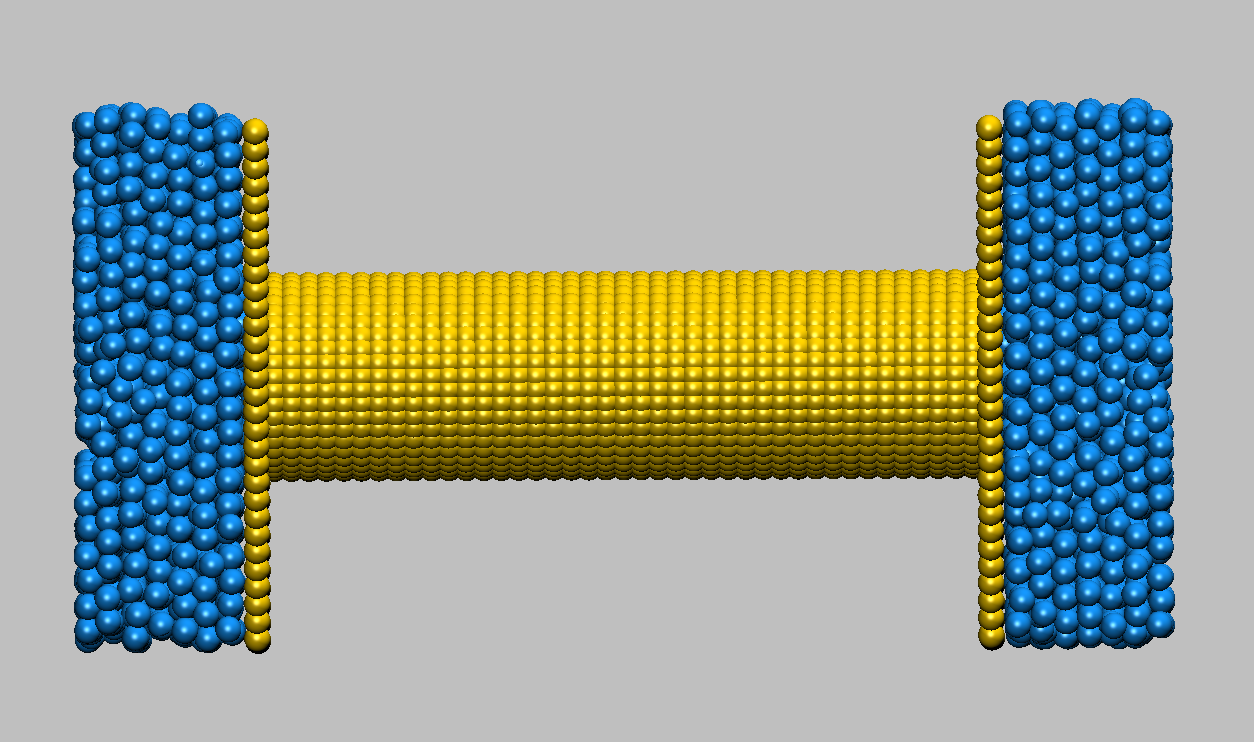} &
\includegraphics[width=.48\columnwidth]{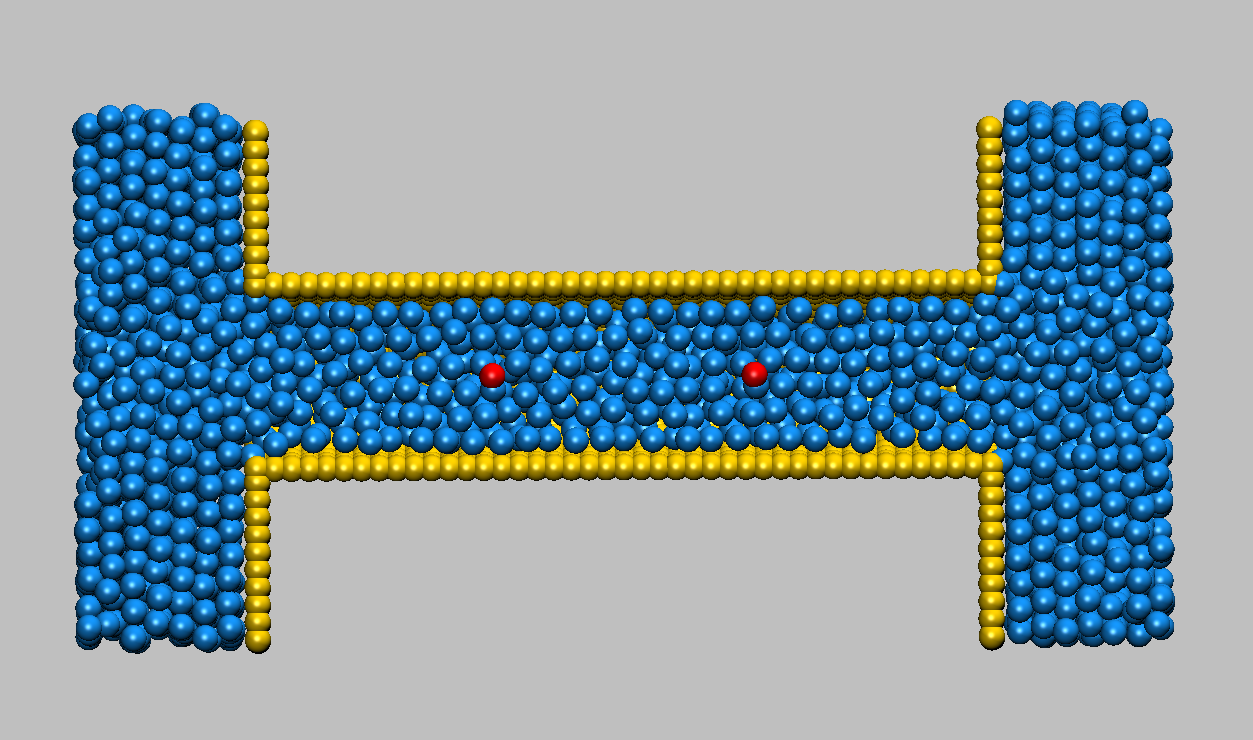}  \\
$(a)$& $(b)$ \\
\end{tabular}
\begin{tabular}{cc}
\includegraphics[width=.48\columnwidth]{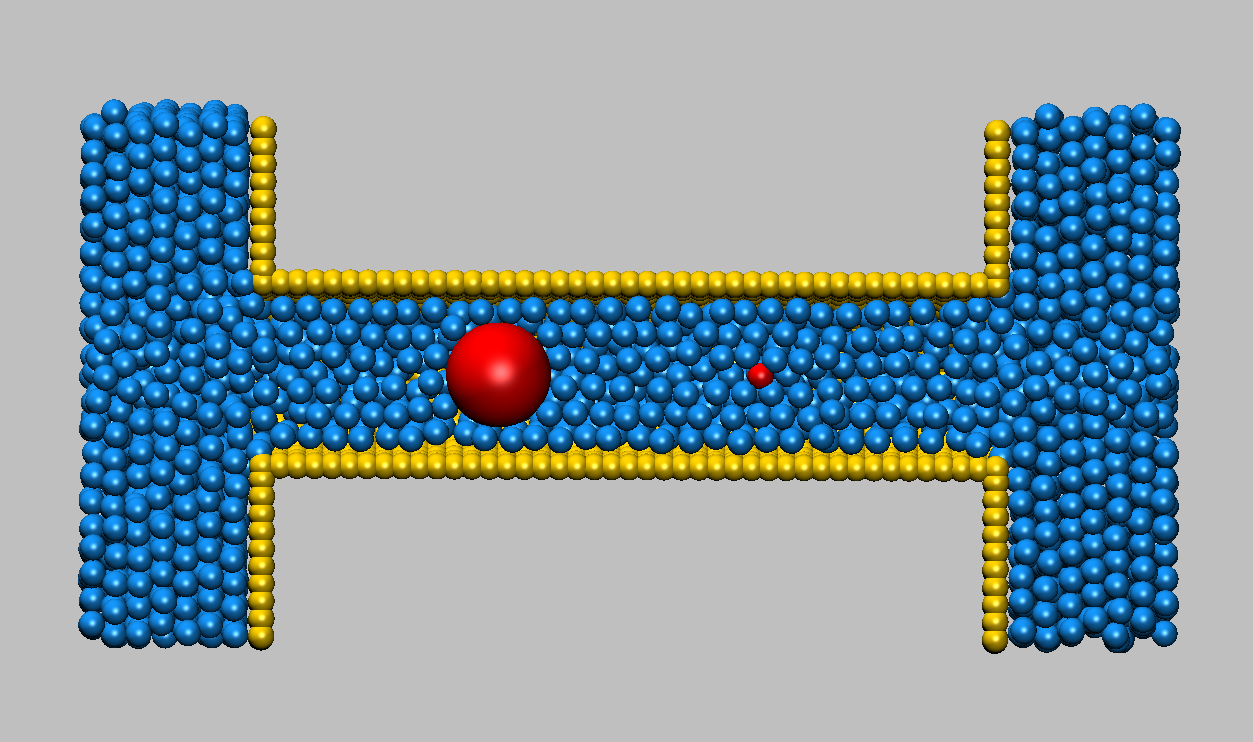} &
\includegraphics[width=.48\columnwidth]{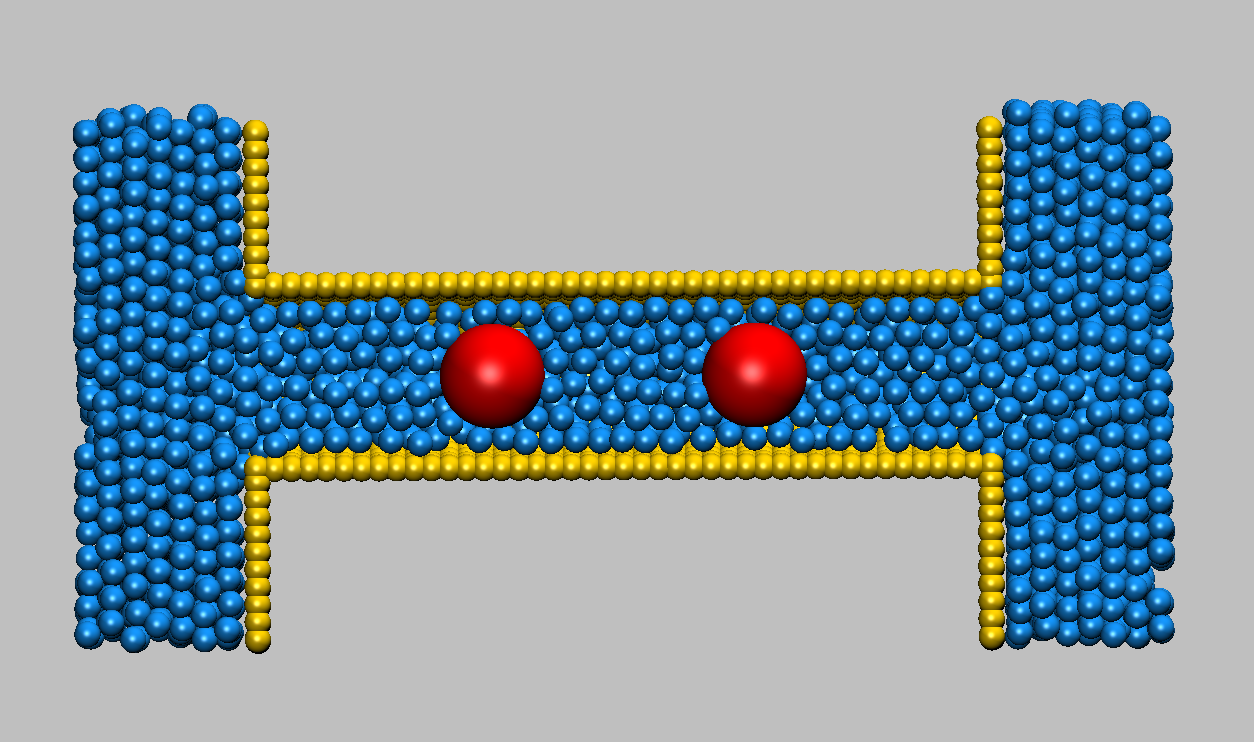}  \\
$(c)$& $(d)$ \\
\end{tabular}
\end{center} \caption{a) Schema of simulation box. b) A look through the tube before starting the stroke. liquid particles (blue) fill the tube and two actuator beads (red) are fixed. c) Right actuator is expanded (part $A1$ is narrowed). d) Both actuators are expanded.
\label{snapshot}}
\end{figure}

\begin{figure}
\begin{center}
\includegraphics[width=1\columnwidth]{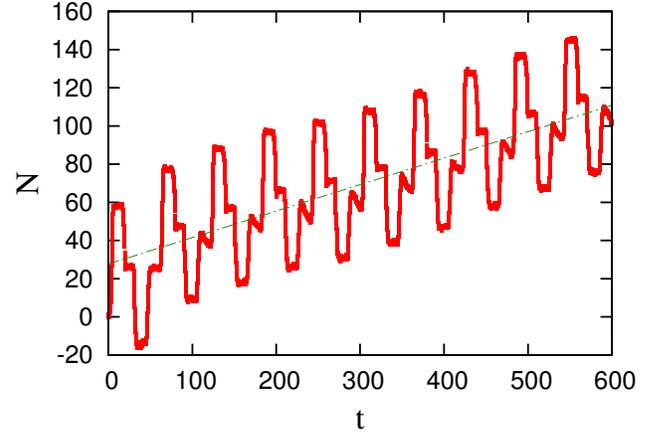}
\end{center} \caption{Number of crossed particles from a cross section of pipe (N) versus time (t) in 10 cycles. Data are avaraged from 20 independent simulations. Dashed line is a linear fit to whole data and can be regarded as flow rate, $Q$. Here $Q=0.14/\tau_0$.\label{profile}}
\end{figure}

\begin{figure}
\begin{center}
\includegraphics[width=1\columnwidth]{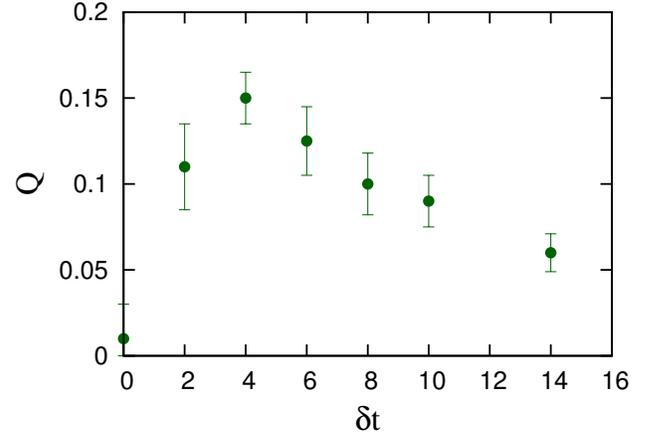}
\end{center} \caption{flow rate versus $\delta t$. Every point in this plot is obtained from 20 MD runs. 
\label{Q}}
\end{figure}

\begin{figure}
\begin{center}
\includegraphics[width=1.0\columnwidth]{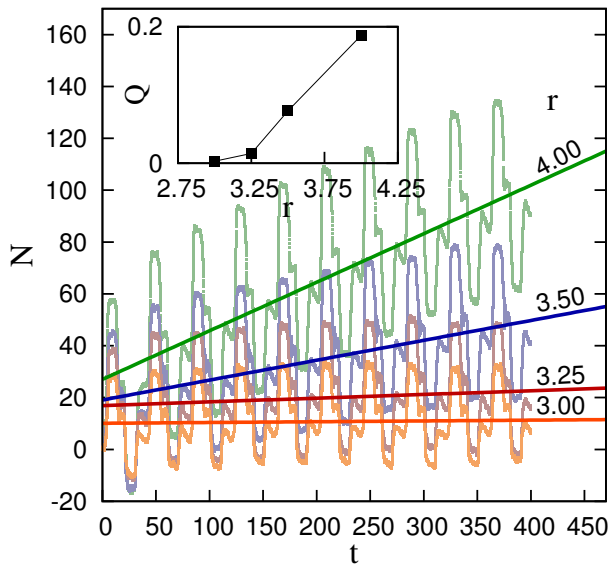}
\end{center} \caption{Number of crossed particles from a cross section of the pipe (N) versus time (t) in 10 cycles for four different value of actuators final radius. Lines are linear fit to data. Each curve is average of 20 different MD runs. Inset shows the dependence of flow rate versus final radius of the actuators. Connecting lines are represented just to guid eyes. 
\label{act_rad}}
\end{figure}

Nanometer pipe is modeled by a fixed carbon nanotube and the fluid filled it, as liquid argon. In Fig. \ref{snapshot} a schema of simulation box included of two reservoirs, connecting pipe, liquid particles and two actuator beads are shown. Interaction potentials between liquid particles and wall, between liquid and two actuators and between liquid particles are Lennard-Jones potential:
\begin{eqnarray}\
V_{LJ}\left(r\right)=4{\epsilon}[{\left(\frac{\sigma}{r}\right)}^{12}-{\left(\frac{\sigma }{r}\right)}^6].
\end{eqnarray}
\\
Lennard-Jones' parameters are fixed as ${\sigma }_{c-Ar}=3.573\ \mathop{\rm A}\limits^\circ$, ${\epsilon}_{c-Ar}=0.2827\ kcal.{mol}^{-1}$, ${\sigma}_{Ar-Ar}=3.35\ \mathop{\rm A}\limits^\circ $  and ${\epsilon}_{Ar-Ar}=0.2862\ kcal.{mol}^{-1}$~\cite{Tuzun1996}. The simulation temperature and density are 80${K}$ and $1.4 g/cm^3$ respectively. From now on we use $Ar$ atom diameter as the unit of length, $\sigma_0$ and $Ar-Ar$ interaction energy as the unit of energy, $\myepsilon_0$ so time would be given in units of $\tau_0=\sigma_0\sqrt{m_0/\myepsilon_0}=1.96\times10^{-12}s$, where $m$ denotes the mass of an Argon atom. Unless explicitly stated, all the quantities in this paper are given in these reduced units. Before actuating the actuators, we let simulation run for 100000 steps to reach equilibrium. The expansion and contraction time of actuators are $5\tau_0$ and the delay time between two consecutive stages, $\delta t$, varies depending on the situation. Simulation box is periodic in all direction and so pressure of two reservoirs are the same.
\\
Fig. \ref{profile} shows a sample flow profile (number of fluid particles crossing a cross section of the pipe to the right) versus time. In these series of simulation $\delta t=10$, tube radius is $4$ and first and final radius of the actuator beads are $1$ and $4$, respectively. This flow profile interestingly follows the actuators' operation and has distinctive rises (falls) at the beginning of each expansion (contraction) stage. 
In Fig. \ref{Q} flow rate (slope of $N-t$ diagram), $Q$, versus $\delta t$, delay time between two consecutive stages, is plotted. Here, tube radius is $3$ and first and final radius of the actuator beads are $0.5$ and $2.5$. At small $\delta t$, fluid has not enough time to flow in true direction and at big $\delta t$, the delay is so long (much bigger than equilibrium time of the system) that fluid particles have enough time to reach their equilibrium state and rearrange again uniformly trough the simulation box and hence pump is not efficient in very small and big delay times. Such a dependence on delay time or frequency is generally observable in peristaltic pumps \cite{Lee2012a,Goldschmidtboing2005}.
\\
Fig. \ref{act_rad} shows the effect of actuator radius on flow process. For four different value of final radius of actuator beads, $r$, flow profile is sketched versus time. Here $\delta t=5$, tube radius is $4$ and first radius of the actuator beads is $1$. As it can be seen the effective pump is the one with complete contraction and flow rate decreases very fast as final radius decreases. As it is expected the maximum flow rate corresponds to full clogging of the tube.
\\

\section{Conclusion}
In this paper it is discussed that the simple designed peristaltic pump of Berg is able to generate an effective flow in a distinctive direction in nano scale. This can be shown by a simple theoretical model in which we consider that the change in the radius of tube yields to the change in flow resistance of two actuator parts of the tube in a non-reciprocal way. For the first time a modest interpretation of NG swimmer is proposed which has the essential features of the system to model its non-reciprocal behavior. It is shown that there is an analogy between NG swimmer and this pump, in both system non reciprocal change of qualitative parameters breaks the symmetry of space and different resistance in two parts of the system makes an effective relative motion between the fluid and the system. In our simple model, Berg pump is composed of a long tube connecting two fluid reservoirs. Two flexible parts, with two natural states, contracted and expanded, are fixed in two points along it. These two flexible parts act as actuators of the pump and by their periodic contraction and expansion, enforce the fluid to flow in a defined direction. In reality these actuators can be two proteins with two natural states, expanded and contracted, which by a specific active control, like emission of an electromagnetic wave, can switch between these two states. Biological gates such as OmpF are other candidates for doing these job. By considering difficulties of experimental design, only as a suggestion, one can consider a square
pulse wave that is propagating from left of the setup (Fig. \ref{nanopump}) and its amplitude oscillates between two value, one of them in the range that can makes the actuator protein to expand and the other forces it to contract. If the wave length of this wave is at least a bit larger than the distance between two actuator parts, process described in Fig. \ref{nanopump} would be simulated.
\\ 
We simulated this nanoscale pump with a long narrow tube and two actuator beads and investigated its behavior under different frequency of actuator's oscillation and tube radius. There is a frequency in which this pump have maximum fluid flow and this frequency depends on
other parameters of the system such as tube radius.
\bibliographystyle{apsrev}
\bibliography{biblio}
\end{document}